\documentstyle[epsfig,here,12pt]{article}

\newcommand{\preprint}[1]{\begin{flushright}#1\end{flushright}}

\newcommand{\bea}{\begin{eqnarray}}
\newcommand{\eea}{\end{eqnarray}}
\newcommand{\simgt}{\hbox{ \raise3pt\hbox to 0pt{$>$}\raise-3pt\hbox{$\sim$} }}
\newcommand{\simlt}{\hbox{ \raise3pt\hbox to 0pt{$<$}\raise-3pt\hbox{$\sim$} }}

\begin{document}

\preprint{TU-626}

\vspace*{2cm}
\begin{center}
{\bf \Large Top quark mass determination near $t\bar{t}$ threshold \\ at lepton colliders 
\footnote{Talk given at the KEK LC meeting, March 15-17, 2001.}}
\\[10mm]
{\large Y. Kiyo}
	        \\[10mm]
{\it Department of Physics, Tohoku University\\
		          Sendai, 980-8578 Japan }
\end{center}

\vspace{0.5cm}

\begin{abstract}
Future $e^+ e^-$ linear colliders will enable us to determine the top quark 
mass with high accuracy from the measurement of the 1S peak position of (remnant of)
toponium. The estimated statistical error in this measurement is about 50 MeV
with integrated luminosity of 30 ${\rm fb^{-1}}$.
We calculate ${\cal O}(\alpha_s^5 m)$ corrections to the quarkonium 1S
energy spectrum in the large-$\beta_0$ approximation to reduce theoretical 
uncertainties below the corresponding experimental error.
We discuss the significance of the ${\cal O}(\alpha_s^5 m)$ corrections 
and estimate theoretical uncertainties of our prediction.

\end{abstract}

\section{Introduction}

At future lepton colliders, we will be able to determine the mass of the
(remnant of) toponium 1S state from a threshold scan of the $t \bar{t}$ 
cross section. The statistical error in this measurement was estimated to be
about $50$ MeV with a moderate integrated luminosity of ${\rm 30 fb^{-1}}$
\cite{PeraltaMartinezMiquel}. 
On the other hand we can predict the mass of toponium 1S state 
as a function of the top quark pole-mass $M_t$ and QCD coupling constant 
$\alpha_s$ in perturbative QCD:
\begin{eqnarray}
M({\rm 1S}(t\bar{t}))
&=& 
2 M_t-\frac{4}{9} \alpha^2_{s} M_t 
+{\cal O}(\alpha_s^3 M_t).
\label{eq:1SMass}
\end{eqnarray}
Using this relation the top quark mass can be extracted from the 
measurement of the 1S peak position of the $t\bar{t}$ total cross section. 

To determine the top quark mass accurately, calculations of the higher order
corrections to Eq.(\ref{eq:1SMass}) are required.
Presently the theoretical prediction for the quarkonium energy spectrum 
\cite{NNLO} is known at next-to-next-to-leading
order (NNLO). Thus we know the relation between the mass of the toponium 1S state
and the top quark pole-mass at ${\cal O}(\alpha_s^4 M_t)$. 
However the series expansion of the toponium mass in $\alpha_s$ is not well-defined 
because there is a large renormalon contribution \cite{AgliettiLigeti}
in the binding energy $E_{bin}$.\footnote{$E_{bin}$ is defined as the energy measured 
from the twice of the quark pole mass, $M({\rm 1S})=2 M_t + E_{bin}$.}
Reliable theoretical predictions can be obtained using the short-distance 
masses. This is because the pole-mass contains the renormalon contribution
\cite{RenormalonPoleMass},
which is cancelled \cite{RenormalonCancellation}
against the renormalon contribution\cite{AgliettiLigeti} in the binding energy. 
After taking account of the renormalon cancellation properly
the perturbative expansion of the toponium energy spectrum shows healthy 
convergence. Use of the pole-mass does not respect the renormalon cancellation,
therefore we should use one of the short-distance masses to obtain a reliable theoretical 
prediction for the quarkonium energy spectrum.  
In this paper we use the $\overline{\rm MS}$-mass  
$\overline{m}_t \equiv m_{t, \overline{\rm MS}}(m_{t, \overline{\rm MS}})$ 
renormalized at the $\overline{\rm MS}$-mass scale. 

Taking renormalon cancellation into account our NNLO prediction of the toponium mass, 
expressed in terms of $\overline{\rm MS}$-mass, has a {\it genuine} 
${\cal O}(\alpha_s^3 \overline{m}_t)$ accuracy which is free from leading renormalon contribution.
The renormalon contribution in ${\cal O}(\alpha_s^3 \overline{m}_t)$ corrections 
to the relation between the pole-mass and the $\overline{\rm MS}$-mass is cancelled 
against the leading renormalon contribution in ${\cal O}(\alpha_s^4 \overline{m}_t)$ 
corrections to the binding energy. Then the remaining  ${\cal O}(\alpha_s^3 \overline{m}_t)$ 
correction to the mass relation determines the accuracy of the present perturbative 
prediction of the toponium mass, which is free from the leading renormalon ambiguity. 
To achieve the top quark $\overline{\rm MS}$-mass determination with 50 
MeV accuracy, the genuine ${\cal O}(\alpha_s^4 \overline{m}_t)$ corrections are
required. 
It is sufficient to calculate further \cite{Sumino00} 
(i) the relation between $\overline{\rm MS}$-mass and pole-mass 
at ${\cal O}(\alpha_s^4 \overline{m}_t)$, and 
(ii) ${\cal O}(\alpha_s^5 \overline{m}_t)$ correction to the 
binding energy in the large-$\beta_0$ approximation, which 
will be sufficient to estimate the leading renormalon contribution.
In Ref.\cite{YY} we have calculated ${\cal O}(\alpha_s^5 m)$ 
correction to quarkonium 1S spectrum in the large-$\beta_0$ 
approximation. In this paper we discuss the significance of this 
${\cal O}(\alpha_s^5 m)$ correction to the binding energy in the top 
quark $\overline{\rm MS}$-mass determination at future lepton colliders. 

The outline of this paper is as follows:
In Sec.2 the relation between the $\overline{\rm MS}$-mass and pole-mass is 
explained. We show the scale dependence of the pole-mass and examine the 
higher order corrections to the top quark pole-mass. We demonstrate that 
the pole-mass is defined only to an accuracy of ${\cal O}(\Lambda_{\rm QCD})$ due 
to the renormalon problem.  
In Sec.3 the renormalon cancellation in the quarkonium energy spectrum is
discussed. We explore the convergence property of the perturbative expansion of 
the toponium 1S mass, and show that we can obtain the reliable perturbative expansion
after taking renormalon cancellation into account.
Sec.4 contains a summary and discussion of the top quark $\overline{\rm MS}$-mass determination
from the measurement of the 1S peak position in $t\bar{t}$ total cross section.

%
\section{Pole-mass and $\overline{\rm MS}$-mass}

In this section we discuss the top quark pole-mass and the renormalon 
in the pole-mass in some details to understand that 
we must renounce the top quark pole-mass for our purpose. 
It is the $\overline{\rm MS}$-mass for which we can reduce the theoretical
uncertainty below ${\cal O}(\Lambda_{\rm QCD})$.

The pole-mass of the quark is defined perturbatively 
as the pole position of the quark propagator. 
The relation between the pole-mass and $\overline{\rm MS}$-mass
is known at three loop \cite{3loopPoleMass}, and higher order 
terms are known in the large-$\beta_0$ approximation \cite{BenekeBraun95}.
Using $\alpha_s(\mu)$ and the $\overline{\rm MS}$-mass\footnote{The relation between 
the pole and $\overline{\rm MS}$ masses has been calculated in the full theory. 
We rewrite this relation using $\alpha_s(\mu)$ which is the coupling constant in the 
"{\it 5-flavor}" effective theory.} the pole-mass of the top quark is written as
\begin{eqnarray}
M_t= \overline{m}_t 
\left\{
1+ C_F \frac{\alpha_s(\mu)}{\pi}
 + C_F \left(\frac{\alpha_s(\mu)}{\pi}\right)^2 d_1
 + C_F \left(\frac{\alpha_s(\mu)}{\pi}\right)^3 d_2
 + C_F \left(\frac{\alpha_s(\mu)}{\pi}\right)^4 d_3
 + \cdots
\right\},
\label{eq:PoleMass}
\end{eqnarray}
where $d_{i}$ are the functions of $\log(\mu/\overline{m}_t)$,
which can be obtained from Refs.\cite{3loopPoleMass,BenekeBraun95}
and running of $\alpha_s(\mu)$. It is known that this perturbative series is divergent due
to the renormalon contribution. The pole-mass has strong sensitivity to soft gluon effects, 
which are the sources of the IR renormalon. The large order behavior of the perturbative 
expansion of the pole-mass can be estimated in the large-$\beta_0$ approximation:
\begin{eqnarray}
d_n
\sim
\frac{\mu}{\overline{m}_t}
\times 
\left(\frac{\beta_0}{2}\right)^n n!,
\label{eq:d_asym}
\end{eqnarray}
where $\beta_0=11-(2/3) n_f$. 
Thus the perturbative expansion is divergent. The formal argument says that the quark 
pole-mass up to all orders, Eq.(\ref{eq:PoleMass}), is scale independent.
But this statement cannot be justified because the series expansion in $\alpha_s$ is not 
very convergent.

\vspace*{0.5cm}
\begin{figure}[h]
\begin{center}
\leavevmode
\psfig{file=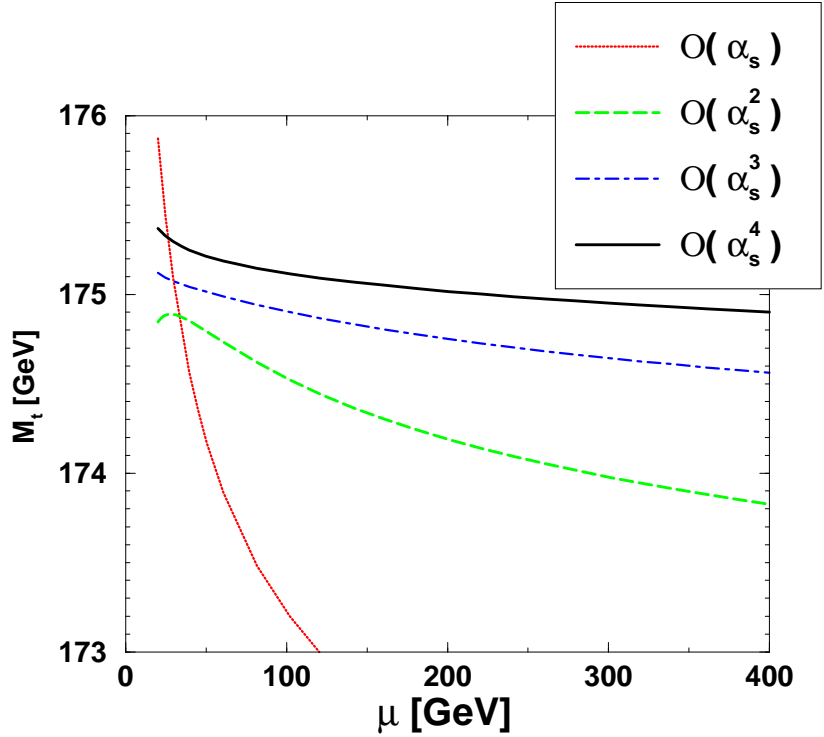,width=10cm,angle=-90}
\caption{Scale dependence of the top quark pole-mass in $\alpha_s$ expansion.
The result in the large-$\beta_0$ approximation is used for ${\cal O}(\alpha_s^4)$ term.
The input parameters are $\alpha_s^{(5)}(M_Z)=0.119$ and 
the $\overline{\rm MS}$-mass of the top quark, $\overline{m}_t=165$ GeV.}
\label{fig:PoleMass}
\end{center}
\end{figure}
%
In Fig.\ref{fig:PoleMass} we show the top quark pole-mass
expressed in terms of $\overline{m}_t$ up to ${\cal O}(\alpha_s^n)~ (n=1,\cdots,4)$ as a 
function of the renormalization scale $\mu$:
\begin{itemize}
\item The scale dependence is large and there is no scale-flat point 
which satisfies the minimum sensitivity condition against $\mu$.\footnote{
We do not take the scale-flat point of ${\cal O}(\alpha_s^2)$ line at $\mu \sim 20$ GeV 
seriously. Indeed this behavior disappears at ${\cal O}(\alpha_s^3)$.} 
Naively the scale-flat point around $\mu \sim {\cal O}(\overline{m}_t)$ is expected
because the only scale in the problem is $\overline{m}_t$. The higher order corrections 
can not rescue the strong $\mu$-dependence of the pole-mass.\footnote{
This is seen more clearly in $\mu$-dependence of the pole-mass 
of the bottom quark, because of large $\alpha_s$ in the series expansion of the 
bottom quark pole-mass.}
\item The higher order term seems to give ${\cal O}(\Lambda_{\rm QCD})$ shift.
This is consistent with the general argument; the ambiguity due
to renormalon contribution is $\delta M_t \sim {\cal O}(\Lambda_{\rm QCD})$ 
\cite{RenormalonPoleMass}.
\end{itemize}

The pole-mass has a strong sensitivity to soft gluon effects, which are the sources 
of the renormalon contribution to the pole-mass. The higher order corrections to 
the pole-mass are large and the scale dependence is strong. The pole-mass
is not well-defined due to the renormalon contribution. In the next section 
we show that the renormalon contribution in pole-mass is cancelled
against the leading renormalon contribution in the binding energy 
when quarkonium energy spectrum is expressed in terms of 
the $\overline{\rm MS}$-mass. 

\section{1S energy spectrum and renormalon}

In this section we study the perturbative expansion of the quarkonium energy 
spectrum, the convergence property and scale dependence in the pole and $\overline{\rm MS}$ 
mass scheme. 
We will see that we can obtain the reliable theoretical prediction of $M({\rm 1S})$ only 
when renormalon cancellation is properly taken into account.
The convergence of the series expansion is nicely improved thanks to renormalon cancellation,
furthermore the scale dependence of $M({\rm 1S})$ is reduced.

To achieve renormalon cancellation, 
we rewrite the binding energy in terms of $\alpha_s(\mu)$ and 
$\overline{\rm MS}$-mass using Eq.(\ref{eq:PoleMass}):
\begin{eqnarray}
E_{bin}&=&
-\frac{\overline{m}_t}{4}
\left(C_F \alpha_s(\mu)\right)^2 
\left\{ 1
 + \frac{\alpha_s(\mu)}{\pi} e_1
 + \left(\frac{\alpha_s(\mu)}{\pi}\right)^2 e_2
 + \left(\frac{\alpha_s(\mu)}{\pi}\right)^3 e_3
 + \cdots 
\right\},
\end{eqnarray}
where $e_i$ are functions of $\log(\mu/C_F \alpha_s(\mu) \overline{m}_t)$.
The renormalon contribution can be seen in the large order behavior 
of $e_i$. Employing the large-$\beta_0$ approximation we obtain
\begin{eqnarray}
e_i \sim \frac{8}{\pi}\left(\frac{\mu}{C_F \alpha_s(\mu) \overline{m}_t} \right) 
\times \left(\frac{\beta_0}{2}\right)^n n!.
\end{eqnarray}
In the total energy $2M_t+E_{bin}$,
the renormalon contribution in $e_i$ is cancelled against the renormalon contribution 
in $d_i$($i=1,2,\cdots$) from $2 M_t$. 
Some comments are in order:
To achieve renormalon cancellation at each order of series expansion 
it is essential to use same coupling $\alpha_s(\mu)$ in the series expansion of binding 
energy $E_{bin}$ and in the relation between the pole-mass and the $\overline{\rm MS}$-mass. 
The renormalon cancellations occur between the terms whose order in $\alpha_s$ differ by one 
\cite{HoangLigetiManohar} in the quarkonium energy spectrum, the origin of extra power of 
$\alpha_s$ is a dynamical scale of the quarkonium system, Bohr radius $ r_B =(C_F \alpha_s M_t)^{-1}$. 
This extra power of $\alpha_s(\mu)$ is compensated by the $1/\alpha_s(\mu)$ in the large order 
behavior of $e_i$.

\begin{figure}[t]
\begin{center}
\hspace*{-2cm}
\leavevmode
\psfig{file=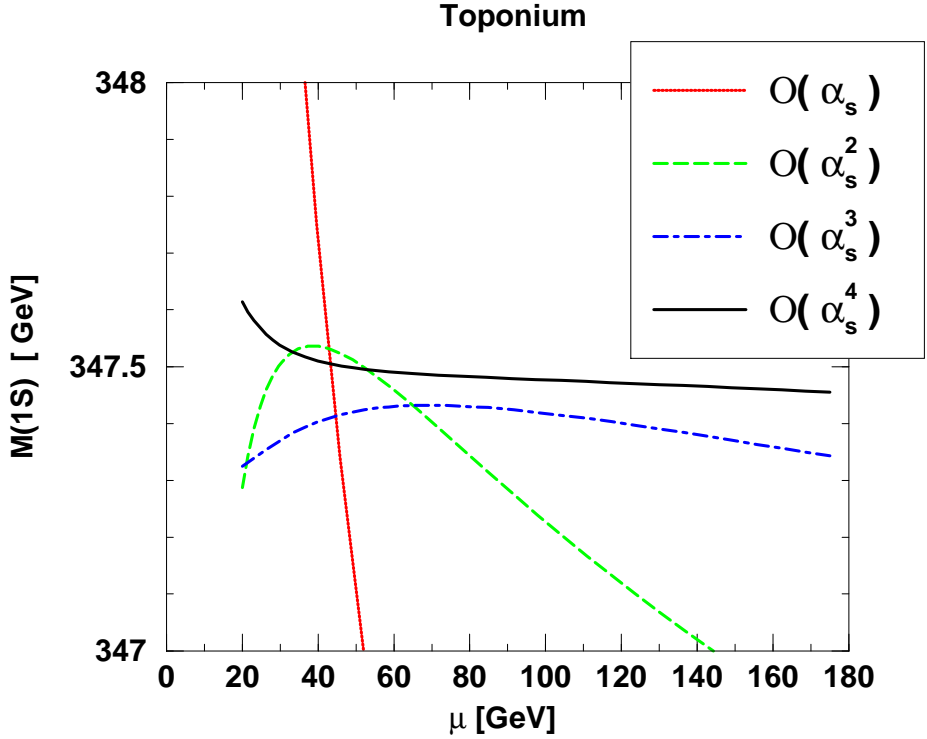,width=10cm,angle=-90}
\caption{Scale dependence of the series expansion of 
$M({\rm 1S})$ in the $\overline{\rm MS}$ scheme. 
The QCD correction to quarkonium energy spectrum begins 
at ${\cal O}(\alpha)$ in the $\overline{\rm MS}$ scheme because 
$M({\rm 1S}) \sim 2 M_t \sim 2 \overline{m} (1+ C_F \alpha_s/\pi)$.
The ${\cal O}(\alpha_s^4)$ term is evaluated in the large-$\beta_0$ approximation.}
\label{fig:M1S_MSbar}
\end{center}
\end{figure}

Now we show our numerical results of $M({\rm 1S})$:
\begin{eqnarray}
M({\rm 1S}) &=&
2 \times 
( 174.79 -  0.77  -  0.35  -  0.25  -  0.13^\star )~{\rm GeV}, ~~(\mbox{pole-mass scheme})
\label{eq:M1S_pole} \\
 &=&
2 \times 
( 165.00 + 7.21 + 1.24 + 0.22 + 0.052^\star )~{\rm GeV}, ~~(\overline{\rm MS}{~\rm scheme})
\label{eq:M1S_MS} 
\end{eqnarray}
where $M_t=174.79$ GeV ($\overline{m}_t=165$ GeV) and $\alpha_s^{(5)}(M_Z)=0.119$ have been used as input
parameters in the pole-mass ($\overline{\rm MS}$) scheme;
the terms with stars are evaluated using the large-$\beta_o$ approximation.
The renormalization scale is taken as $\mu=r_B^{-1} ~(\mu=\overline{m}_t)$ in the 
pole-mass ($\overline{\rm MS}$) scheme. The convergence of the perturbative 
expansion is very slow in the pole-mass scheme, while we see the healthy convergence    
in the $\overline{\rm MS}$ scheme. 
In Figs.\ref{fig:M1S_pole} and \ref{fig:M1S_MSbar}, we show the scale 
dependence of the perturbative expansion of $M({\rm 1S})$ in the pole-mass 
and $\overline{\rm MS}$ scheme, respectively:
\begin{figure}[t]
\begin{center}
\leavevmode
\psfig{file=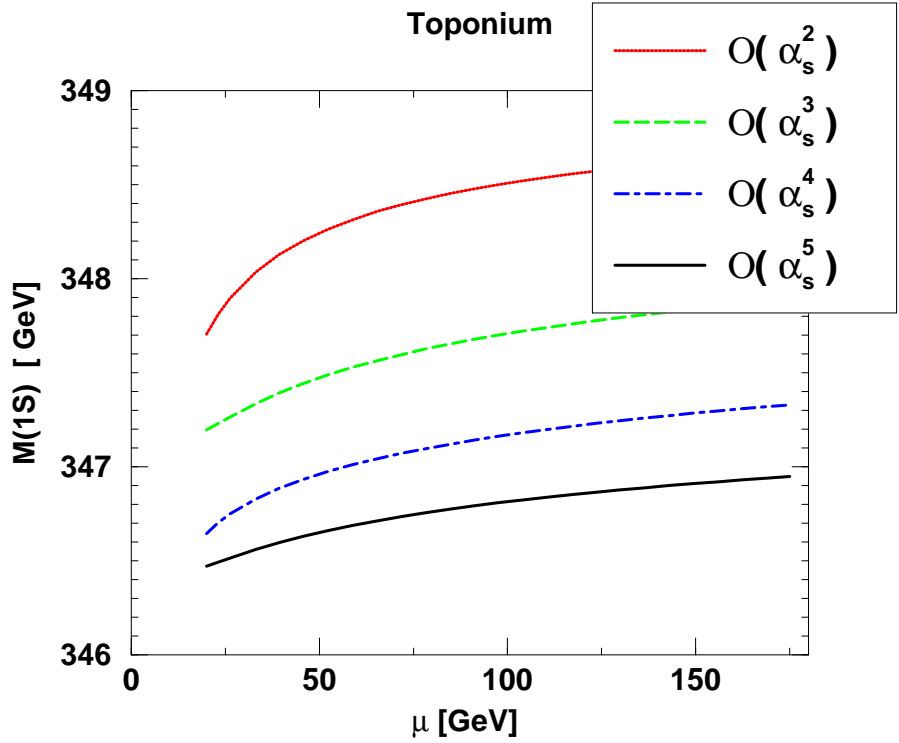,width=10cm,angle=-90}
\caption{Scale dependence of $M({\rm 1S})$ at LO, N$^n$LO 
($n=1,2,3$) are shown in the pole-mass scheme.  
The QCD correction to quarkonium energy spectrum begins at 
${\cal O}(\alpha^2)$ in the pole-mass scheme because 
$E_{bin}\sim \langle C_F \alpha_s/r\rangle\sim C_F^2 \alpha_s^2 M_t$.
The result in the large-$\beta_0$ approximation is used for the 
line of ${\cal O}(\alpha_s^5)$.}
\label{fig:M1S_pole}
\end{center}
\end{figure}
%
\begin{itemize}
\item Scale dependence is strong in the pole-mass scheme, while the $\mu$-dependence is 
very weak in the $\overline{\rm MS}$ scheme. The scale-flat point appears in the 
$\overline{\rm MS}$ scheme at a certain $\mu$ which lies in the region between $1/r_B$ 
and $\overline{m}_t$. This is because we have two large-logs in the $\overline{\rm MS}$ 
scheme, $\log(\mu/\overline{m}_t)$ and $\log(\mu r_B)$
\footnote{A physical interpretation and consequences of this fact are discussed in 
detail for the bottomonium system in Ref.\cite{BrambillaSuminoVairo}.}.

\item In the pole-mass scheme the higher order terms give large corrections despite 
of the smallness of the QCD coupling constant, $\alpha_s \sim 0.1$, while in the 
$\overline{\rm MS}$ scheme we see the healthy convergence of the perturbative expansion.
In the pole-mass scheme the difference between ${\cal O}(\alpha_s^n)$ and ${\cal O}(\alpha_s^{n+1})$ 
lines is an almost constant shift by ${\cal O}(\Lambda_{QCD})\sim 300$ MeV,
which is consistent with the general argument that the ambiguity from the 
renormalon contribution is ${\cal O}(\Lambda_{QCD})$ 
\cite{AgliettiLigeti,RenormalonCancellation}. 
\end{itemize}

\section{Conclusion and Discussion}

We have discussed the relation between the pole-mass and the $\overline{\rm MS}$-mass 
of top quark. The pole-mass always accompanies 
the renormalon ambiguity of ${\cal O}(\Lambda_{\rm QCD})$, thus it is not adequate for the 
precision determination of the top quark mass. 

The relation between the top quark $\overline{\rm MS}$-mass and the toponium mass 
has been investigated at ${\cal O}(\alpha_s^4 \overline{m}_t)$. It is essential to use 
the $\overline{\rm MS}$-mass to obtain a prediction more accurately than the 
leading renormalon ambiguity of ${\cal O}(\Lambda_{\rm QCD})$.
Our theoretical prediction is given at ${\cal O}(\alpha_s^4 \overline{m}_t)$ 
using the large-$\beta_0$ approximation for ${\cal O}(\alpha_s^4 \overline{m}_t)$ 
term. To achieve our goal \cite{sumino01} ultimately, the 4-loop relation between pole and 
$\overline{\rm MS}$ masses should be calculated. 

Finally let us give a discussion on the uncertainty in the relation  between 
$\overline{m}_t$ and $M({\rm 1S})$ given in Eq.(\ref{eq:M1S_MS}). 
The last term in the series expansion of $M({\rm 1S})$ in 
Eq.(\ref{eq:M1S_MS}) is 52 MeV, from which we expect that the uncertainty 
in the top quark $\overline{\rm MS}$-mass determination is below 50 MeV.
This is supported from the scale dependence of mass of toponium 1S state up to 
${\cal O}(\alpha_s^4 \overline{m}_t)$ in Fig.\ref{fig:M1S_MSbar}. 
The last term of ${\cal O}(\alpha_s^4 \overline{m}_t)$ correction in the large-$\beta_0$ 
approximation would be a reasonable estimate of the exact 
${\cal O}(\alpha_s^4 \overline{m}_t)$ term.

\vspace{1cm}
{\bf Acknowledgment:}
Y. K. is grateful to fruitful discussion with Y. Sumino. 
Y.K. was supported by the Japan Society for the Promotion of Science.

%

\end{document}